\documentclass[aps,prl,twocolumn,superscriptaddress,showpacs,letter]{revtex4}
\usepackage{color}
\usepackage{graphicx}
\usepackage{dcolumn}
\usepackage{bm}

\begin{document}

\title{Topological Quantum Computation by Manipulating Quantum Tunneling Effect of the Toric Codes}
\author{Su-Peng Kou}

\affiliation{Department of Physics, Beijing Normal University,
Beijing 100875, China}
\begin{abstract}
Quantum computers are predicted to utilize quantum states to perform
memory and to process tasks far faster than those of conventional
classical computers. In this paper we show a new road towards
building fault tolerance quantum computer by tuning quantum
tunneling effect of the degenerate
quantum states in topological order, instead of by braiding anyons. {%
Using a designer Hamiltonian - the Wen-Plaquette model as an
example, we study its quantum tunneling effect of the toric codes
and show how to control the toric code to realize topological
quantum computation (TQC). In particular, we give a proposal to the
measurement of TQC.} In the end the realization of the Wen-Plaquette
model in cold atoms is discussed.


\end{abstract}

\maketitle

Quantum computers are predicted to utilize quantum states to perform memory
and to process tasks far faster than those of conventional classical
computers. Various designs have been proposed to establish a quantum
computer, including manipulating electrons in a quantum dot or phonon by
using ion traps, cavity QED, or nuclear spin by NMR techniques. {Recently,
people find that it may be possible to incorporate intrinsic fault tolerance
into a quantum computer - topological quantum computation (TQC) which} has
the debilitating effects of decoherence and free from errors. The key point
is to store and manipulate quantum information in a ``non-local'' way, as
means that the ``non-local'' properties of a quantum system remain unchanged
when one does local operations on it. An interesting idea to realize
fault-tolerant quantum computation is anyon-braiding proposed by Kitaev\cite
{k1,k2}. He pointed out that the degenerate ground states of a topological
order make up a protected code subspace (the topological qubit) free from
errors\cite{wen,ioffe}.

It is known that there are two types of topological orders in two dimensions
spin models - non-Abelian topologically ordered state and $Z_2$
topologically ordered state. $Z_2$ topological order is the simplest
topologically ordered state with three types of quasiparticles: $Z_2$
charge, $Z_2$ vortex, and fermions\cite{wen2}. $Z_2$ charge and $Z_2$ vortex
are all bosons with mutual $\pi $ statistics between them. The fermions can
be regarded as bound states of a $Z_2$ charge and a $Z_2$ vortex. Recently,
several two dimensional exactly solved spin models with $Z_2$ topological
orders were found, such as the toric-code model \cite{k1}, the Wen's
plaquette model \cite{wen,wen1} and the Kitaev model on a honeycomb lattice
\cite{k2}. One can operate on the protected code subspace of $Z_2$
topological order by creating an excitation pair, moving one of the
excitations around the torus, and annihilating it with the other. However,
the non-local operations do not form a complete basis.

On the other hand, in non-Abelian topological orders the elementary
excitation becomes non-Abelian anyon with nontrivial statistics. Now
people focus on realizing TQC by braiding non-Abelian anyons. The
degenerate states undergo a nontrivial unitary transformation when a
non-Abelian anyon moves around the other. One can initialize,
manipulate and measure the degenerate ground states with several
non-Abelian anyons\cite{k2,sa1}. Along the road, the key point is to
manipulate single quasi-particle which becomes a hot issue recently
\cite{pa,sarma,du1,zoller,vid,zhang,zhang1,gao}. In particular, the
idea of creation and manipulation of anyons based on controlled
string operators is proposed\cite{zoller,vid}.

Kitaev have noted the degenerate ground states (on a torus) as \emph{the
toric code}. In this paper We will design TQC by manipulating the degenerate
ground states of $Z_2$ topological orders (the toric code), instead of that
by braiding anyons in a non-Abelian topological order. The key point to
manipulate the degenerate ground states is to tune quantum tunneling effect
by controlling external field on spin models. Firstly the effective theory
of the toric code in $Z_2$ topological orders is formalized. Secondly, by
using the Wen-Plaquette model as an example, we show how to control the
toric code by tuning the tunneling of the degenerate ground states\cite
{wen,wen1,wen2}. In this part, we will concentrate on the measurement of
toric code. Finally we give a short discussion on the realization of the
Wen-Plaquette model in cold atoms.

\textit{The effective model of the toric code }: For $Z_2$ topological
orders on an even-by-even ($e*e$) lattice on a torus, the ground state
degeneracy is $4$\cite{wen,wen1,wen2}$.$ The four degenerate ground states
are denoted by $\mid m,$ $n\rangle ,$ $m=0,$ $1$ and $n=0,$ $1$. $|m,$ $%
n\rangle $ have different boundary conditions of fermion's wave-functions $%
\psi (x,$ $y)$ as
\begin{equation}
\psi (x,\text{ }y)=(-1)^m\psi (x,\text{ }y+L_y),\text{ }\psi (x,\text{ }%
y)=(-1)^n\psi (x+L_x,\text{ }y).
\end{equation}
Physically, topological degeneracy arises from presence or the absence of $%
\pi $ flux of fermions through the two holes of the torus. The degenerate
ground states make up two qubits which can be mapped onto quantum states of
two pseudo-spins $\mathbf{\hat{\tau}}_1$ and $\mathbf{\hat{\tau}}_2,$ $\mid
0,$ $0\rangle \rightarrow \mid \uparrow \rangle _1\otimes \mid \uparrow
\rangle _2,$ $\mid 1,$ $0\rangle \rightarrow \mid \downarrow \rangle
_1\otimes \mid \uparrow \rangle _2,$ $\mid 0,$ $1\rangle \rightarrow \mid
\uparrow \rangle _1\otimes \mid \downarrow \rangle _2,$ $\mid 1,$ $1\rangle
\rightarrow \mid \downarrow \rangle _1\otimes \mid \downarrow \rangle _2.$

It is known that the degenerate ground states with topological orders have
same energy in thermodynamic limit. In a finite system,$\ $the degeneracy of
the ground states is (partially) removed due to tunneling processes, of
which a virtual quasi-particle moves around the torus before annihilated
with the other one\cite{k1,wen,ioffe}. In $Z_2$ topological orders, there
are nine tunneling processes denoted by $\mathcal{C}_v^x$, $\mathcal{C}_v^y,$
$\mathcal{C}_v^{x+y},$ $\mathcal{C}_f^x$, $\mathcal{C}_f^y,$ $\mathcal{C}%
_f^{x+y},$ $\mathcal{C}_c^x$, $\mathcal{C}_c^y,$ $\mathcal{C}_c^{x+y},$ that
correspond to virtual $Z_2$-vortex, fermion, $Z_2$ charge propagating along $%
\hat{e}_x$, $\hat{e}_y$ and $\hat{e}_x+\hat{e}_y$ (or $\hat{e}_x-\hat{e}_y$)
directions around the torus, respectively. For example, the process $%
\mathcal{C}_v^x$ becomes the unitary operation as
\begin{equation}
(
\begin{array}{llll}
\mid \uparrow \rangle _1, & \mid \downarrow \rangle _1, & \mid \uparrow
\rangle _2, & \mid \downarrow \rangle _2
\end{array}
)\rightarrow (
\begin{array}{llll}
\mid \downarrow \rangle _1, & \mid \uparrow \rangle _1, & \mid \uparrow
\rangle _2, & \mid \downarrow \rangle _2
\end{array}
).
\end{equation}
Hence we can use the a pseudo-spin operator $\tau _1^x\otimes \mathbf{1}$ to
represent $\mathcal{C}_v^x$. Similarly $\mathcal{C}_v^y,$ $\mathcal{C}%
_v^{x+y},$ $\mathcal{C}_c^x$, $\mathcal{C}_c^y,$ $\mathcal{C}_c^{x+y}$, $%
\mathcal{C}_f^x,$ $\mathcal{C}_f^y,$ $\mathcal{C}_f^{x+y}$ can be denoted by
$\mathbf{1}\otimes \tau _2^x$, $\tau _1^x\otimes \tau _2^x$, $\tau
_1^x\otimes \tau _2^z,$ $\tau _1^z\otimes \tau _2^x$, $\tau _1^y\otimes \tau
_2^y,$ $\mathbf{1}\otimes \tau _2^z,$ $\tau _1^z\otimes \mathbf{1}$, $\tau
_1^z\otimes \tau _2^z,$ respectively. Among them, one can choose four basic
processes, $\mathcal{C}_v^x$, $\mathcal{C}_v^y,$ $\mathcal{C}_f^x$, $%
\mathcal{C}_f^y,$ and check the ground state degeneracy from the commutation
relations between $\mathcal{C}_v^x$ and $\mathcal{C}_f^y$ , that obeys the
Heisenberg algebra$,$ $\mathcal{C}_v^x\mathcal{C}_f^y=e^{i\pi }\mathcal{C}%
_f^y\mathcal{C}_v^x.$

Hence the effective Hamiltonian of the toric code is obtained in term of the
pseudo-spin operators :
\begin{widetext}
\begin{equation}
\mathcal{H}_{\mathrm{eff}}=J_{xx}\tau _1^x\mathbf{\cdot }\tau
_2^x+J_{yy}\tau _1^y\mathbf{\cdot }\tau _2^y+J_{zz}\tau _1^z\mathbf{\cdot }%
\tau _2^z+J_{zx}\tau _1^z\mathbf{\cdot }\tau _2^x+J_{xz}\tau _1^x\mathbf{%
\cdot }\tau _2^z+\tilde{h}_1^x\tau _1^x+\tilde{h}_1^z\tau _1^z+\tilde{h}%
_2^x\tau _2^x+\tilde{h}_2^z\tau _2^z{\mathbf{\cdot }}  \nonumber
\end{equation}
\end{widetext}
Here $J_{xx},$ $J_{yy}$, $J_{zz},$ $J_{zx}$, $J_{xz},$ $\tilde{h}_1^x,$ $%
\tilde{h}_2^x$, $\tilde{h}_1^z,$ $\tilde{h}_2^z$ are determined by the
energy splitting of the degenerate ground states from the nine tunneling
processes in a $Z_2$ topological order$.$

\textit{Manipulating the toric code by controlling tunneling splitting : }To
design a topological quantum computer, one needs to do arbitrary unitary
operations on the toric code. To emphasize this point, we introduce a
concept '\emph{Controllable Topological Order (CTO)}'. In a CTO,
quasi-particles' dispersions and the energy splitting of the degenerate
ground states can be manipulated\cite{zoller,vid}.

In the following part, based on an example of controllable $Z_2$ topological
order\emph{\ }- the Wen-plaquette model, we demonstrate how to do TQC. The
Hamiltonian of the Wen-plaquette model is\cite{wen,wen1,wen2}
\begin{equation}
H_0=-{g\sum_iF_i}.
\end{equation}
Here ${F_i=\sigma _i^x\sigma _{i+\hat{e}_x}^y\sigma _{i+\hat{e}_x+\hat{e}%
_y}^x\sigma _{i+\hat{e}_y}^y}$ and $g>0.$ ${\mathbf{\sigma }_i}$ are Pauli
matrices on sites, $i.$ The ground state denoted by ${F_i\equiv +1}$ at each
site is a $Z_2$ topological state with the ground state energy $E_g=-g%
\mathcal{N}$ where $\mathcal{N}$ is the total lattice number\cite{wen2}. On
an $e*e$ lattice, the four degenerate ground states make up two qubits.
However, on even-by-odd ($e*o$), odd-by-even ($o*e$) and odd-by-odd ($o*o$)
lattices, the ground state degeneracy is two instead of four$.$

In this model $Z_2$ charge is defined as ${F_{i\in even}=-1}$ at even
sub-plaquette and $Z_2$ vortex is ${F_{j\in odd}=-1}$ at odd sub-plaquette.
The fermions can be regarded as bound states of a $Z_2$ charge and a $Z_2$
vortex on two neighbor plaquettes. These quasi-particles in such exactly
solved model have flat band. In other words, the quasiparticles cannot move
at all. Under the perturbation
\begin{equation}
H^{\prime }=h^x\sum\limits_i\sigma _i^x+h^z\sum\limits_i\sigma _i^z,
\end{equation}
the quasiparticles begin to hop. The term $h^x\sum\limits_i\sigma _i^x$
drives the $Z_2$ vortex, $Z_2$ charge and fermion hopping along diagonal
direction $\hat{e}_x+\hat{e}_y$. The term $h^z\sum\limits_i\sigma _i^z$
drive fermion hopping along $\hat{e}_x$ and $\hat{e}_y$ directions without
affecting $Z_2$ vortex and $Z_2$ charge. One can see the detailed
description of the three kinds of quasi-particles in Ref.\cite{wen2}.

As a result, there exist five tunneling processes under the perturbation $%
H^{\prime },$ $\mathcal{C}_v^{x+y},$ $\mathcal{C}_f^{x+y},$ $\mathcal{C}_f^x$%
, $\mathcal{C}_f^y,$ $\mathcal{C}_c^{x+y}.$ Let us calculate the ground
state energy splitting\ from a higher\ order (degenerate) perturbation
approach. From the tunneling processes of $Z_2$ vortex and $Z_2$ charge\cite
{k1,wen,ioffe}, $\mathcal{C}_v^{x+y}$ and $\mathcal{C}_c^{x+y}$, one can
determine $J_{xx}=J_{yy}=J\sim g(\frac{h_x}g)^L.$ For a $L_x\times L_y$
lattice on a torus, $L$ is equal to $\frac{L_xL_y}\xi $ where $\xi $ is the
maximum common divisor for $L_x$ and $L_y$. From the tunneling process of
fermion, $\mathcal{C}_f^x,$ $\mathcal{C}_f^y$ and $\mathcal{C}_f^{x+y}$, one
can obtain $\tilde{h}_1^z\sim g(\frac{h_z}{2g})^{L_y}$, $\tilde{h}_2^z\sim g(%
\frac{h_z}{2g})^{L_x}$ and $J_{zz}\sim g(\frac{h_x}g)^{2L}+g(\frac{h_z}{2g}%
)^{L_x+L_y}.$\ Other parameters are all zero, $J_{zx}=J_{xz}=\tilde{h}_1^x=%
\tilde{h}_2^x=0$. Then the effective model is simplified into

\begin{equation}
\mathcal{H}_{\mathrm{eff}}\simeq J(\tau _1^x\mathbf{\cdot }\tau _2^x+\tau
_1^y\mathbf{\cdot }\tau _2^y)+J_{zz}\tau _1^z\mathbf{\cdot }\tau _2^z++%
\tilde{h}_1^z\tau _1^z+\tilde{h}_2^z\tau _2^z{\mathbf{\cdot }}  \nonumber
\end{equation}

On $e*o$, $o*e$ or $o*o$ lattices, the effective model $\mathcal{H}_{\mathrm{%
eff}}$ becomes more simple. For example, the two degenerate ground states on
an $L*L$ $($ $L$ is an odd number$)$ lattice are $\mid \uparrow \rangle
_1\otimes \mid \uparrow \rangle _2$ and $\mid \downarrow \rangle _1\otimes
\mid \downarrow \rangle _2$ which are denoted as $\mid \uparrow \rangle $
and $\mid \downarrow \rangle $ in the following parts of this paper. And the
pseudo-spin operator from different tunneling processes is denoted by $%
\mathbf{\tau }$. Then the dynamics of a single qubit can be described by a
simple effective pseudo-spin Hamiltonian
\begin{equation}
\mathcal{H}_{\mathrm{eff}}=J(\tau ^x+\tau ^y)+J_{zz}\tau ^z.
\end{equation}
Fig.1 shows the energy splitting (scaled by $g$ ) of the two degenerate
ground states from the exact diagonalization numerical results.

Then by adding the specific perturbations to the Wen-plaquette model, $%
H^{\prime },$ one can change different quasi-particles' hopping and then
manipulate the toric code by controlling tunneling splitting of degenerate
ground states. Such dramatic property in the Wen-plaquette model gives an
example of so-called \emph{controllable topological orders}. Similar
properties have been used to creat and manipulate anyons in the Kitaev model%
\cite{zoller,vid}.

\textit{Topological quantum computation : }In this section we focus on two
degenerate ground states on an $o*o$ lattice ($\mid \uparrow \rangle $ and $%
\mid \downarrow \rangle $) and show the initialization, the unitary
transformation and the measurement.

\begin{figure}[tbp]
\includegraphics[width=0.4\textwidth]{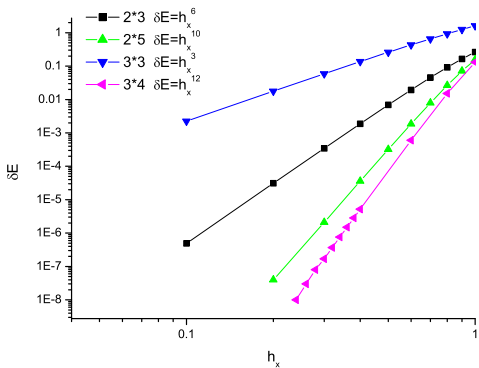}
\caption{The energy splitting between the two degenerate ground states $%
\delta E$ from the exact diagonalization numerical results (g is set to be
unit, g=1). Here $N*M$ denotes a $N\times M$ lattice.}
\label{Fig.1}
\end{figure}

Firstly we initialize the system into the quantum state $\mid \uparrow
\rangle $. This process will occur according to the Hamiltonian $H^{\prime }{%
=h(t)\sum\limits_i\mathbf{\sigma }_z}$ where ${h(t)=e}^{-t/t_0}-1$. At first
$t\rightarrow -\infty $, the ground state is the spin-polarized state of $%
\sigma ^z$ eigenstates $|{\psi _0}\rangle $. Since the effective Hamiltonian
of the toric code is $\mathcal{H}_{\mathrm{eff}}=\tilde{h}^z\cdot \tau ^z,$
the time-evolution operator $U(t)=e^{\frac{-iH^{\prime }t}\hbar }$ in the
topologically ordered phase becomes a projection operator of the pseudo-spin
as $U(t)|{\psi }\rangle \rightarrow \mid \uparrow \rangle $. The system will
eventually evolve \emph{adiabatically and continuously} from $|{\psi _0}%
\rangle $ into the final state $U(t=0)|{\psi _0}\rangle \rightarrow \mid
\uparrow \rangle $. $\mid \uparrow \rangle $ becomes the initial state
prepared for TQC.

Secondly we show unitary operations of the TQC\cite{du1,zhang,zoller}. A
general pseudo-spin rotation operator is defined as
\begin{equation}
U_{\theta ,\varphi }=e^{-\frac i\hbar \gamma \tau ^z}e^{-\frac i\hbar
\varphi (\tau ^x+\tau ^y)}e^{-\frac i\hbar \theta \tau ^z}
\end{equation}
where $\gamma =J_{zz}\Delta {t}_\gamma ,$ $\theta =J_{zz}\Delta {t}_\theta $
and $\varphi =J\Delta {t}_\varphi .$ For example, the Hadamard gate can be a
special pseudo-spin rotation operator, $U_{\theta ,\varphi }(\gamma {=\frac
\pi 4,}$ $\theta {=\frac{7\pi }4,}$ $\varphi =\frac \pi 4).$ To design the
Hadamard gate, we may apply the external field along \textrm{z}-direction at
an interval $\Delta {t}_\theta ={\frac{7\pi }{4J_{zz}}}$ {firstly}${.}$
Then, we swerve the external field along \textrm{x}-direction at an interval
$\Delta {t}_\varphi ={\frac \pi {4J}.}$ Finally, the external field along
\textrm{z}-direction is added at an interval $\Delta {t}_\gamma ={\frac \pi
{4{J_{zz}}}.}$ Using similar method, one can reach certain quantum
operations demanded by TQC and have the ability to carry out arbitrary gate
onto the toric code at will, $\alpha \mid \uparrow \rangle +\beta e^{i\phi
}\mid \downarrow \rangle $ with $\alpha ,$ $\beta \geq 0$ ($\alpha ^2+\beta
^2=1$).

Thirdly let us discuss the measurement of an arbitrary state $\mid \mathrm{%
vac}\rangle =\alpha \mid \uparrow \rangle +\beta e^{i\phi }\mid \downarrow
\rangle $. The interference from Aharonov--Bohm (AB) effect allows one to
design an experimentally observable distinction between the processes with
or without a $\pi $-flux inside the loop. To determine $\alpha ,$ $\beta $
and $\phi ,$ we need to observe both fermion interference and $Z_2$ vortex
interference. Fig.2 is a scheme to show the AB interference on a torus.

Let's explain how to determine $\alpha $ and $\beta $ by AB effect from
fermion-interference. To observe the AB interference, we add a small
external field $h^x\rightarrow 0$ and $h^z=0$. Now $Z_2$ vortex, $Z_2$
charge and fermion hop along diagonal direction $\hat{e}_x+\hat{e}_y.$ On a
torus, there exist two symmetrical paths between two sites ( $i$ and $j$ )
on opposite positions on a torus, $\gamma _1$ and $\gamma _2$. Then the two
trajectories will contribute to the transition amplitude $T_{i,j}$ according
to :
\begin{equation}
T_{i,j}={\left| \Psi _{i,j}^{\gamma _1}\right| ^2}+{\left| \Psi
_{i,j}^{\gamma _2}\right| ^2}+2\epsilon \left| \Psi _{i,j}^{\gamma _2}\Psi
_{i,j}^{\gamma _1}\right|
\end{equation}
where $\Psi _{i,j}^{\gamma _2}$ and $\Psi _{i,j}^{\gamma _1}$ are the wave
functions of fermions of the two trajectories. For the ground state $\mid
\uparrow \rangle $, $\epsilon $ is unit, $\epsilon =1$. However, for the
ground state $\mid \downarrow \rangle,$ $\epsilon =-1.$ Then we can
distinguish these two cases. For two symmetrical paths $\Psi _{i,j}^{\gamma
_2}=\Psi _{i,j}^{\gamma _1}{=t}_f$, we get a probability $\alpha ^2$ for $%
\mid \uparrow \rangle $ with $T_{i,j}=4{t_f^2}$ and a probability $\beta ^2$
for $\mid \downarrow \rangle $ with $T_{i,j}=0$.

On the other hand, one can determine the parameter $\phi $ by observing $Z_2$
vortex interference from $i$ to site $j$. The wave function of $Z_2$ vortex
has a periodic boundary condition along $x$ direction for the ground state $%
\mid \uparrow ^{\prime }\rangle =\frac 1{\sqrt{2}}\mid \uparrow \rangle
+\mid \downarrow \rangle $ and an anti-periodic boundary condition for the
ground state $\mid \downarrow ^{\prime }\rangle =\frac 1{\sqrt{2}}\mid
\uparrow \rangle -\mid \downarrow \rangle .$ Then an arbitrary state $\mid
\mathrm{vac}\rangle =\alpha \mid \uparrow \rangle +\beta e^{i\phi }\mid
\downarrow \rangle $ is re-written into
\begin{equation}
\mid \mathrm{vac}\rangle =\sqrt{\frac 12+\alpha \beta \cos \phi }e^{i\phi
^{\prime }}\mid \uparrow ^{\prime }\rangle +\sqrt{\frac 12-\alpha \beta \cos
\phi }\beta e^{i\phi ^{\prime \prime }}\mid \downarrow ^{\prime }\rangle
\end{equation}
where $\phi ^{\prime }=\arctan (\frac{\sin \phi }{\beta \cos \phi +\alpha })$
and $\phi ^{\prime \prime }=\arctan (\frac{\sin \phi }{\beta \cos \phi
-\alpha })$. For two symmetrical paths $\Psi _{i,j}^{\gamma _2}=\Psi
_{i,j}^{\gamma _1}{=t}_v$, we get a probability $(\frac 12+\alpha \beta \cos
\phi )$ for $\mid \uparrow ^{\prime }\rangle $ with $T_{i,j}=4{t_v^2}$ and a
probability $\frac 12-\alpha \beta \cos \phi $ for $\mid \downarrow ^{\prime
}\rangle $ with $T_{i,j}=0$. As a result, we determine the parameters $%
\alpha ,$ $\beta $ and $\phi $ of an arbitrary state $\mid \mathrm{vac}%
\rangle =\alpha \mid \uparrow \rangle +\beta e^{i\phi }\mid \downarrow
\rangle $.

\begin{figure}[tbp]
\includegraphics[clip,width=0.4\textwidth]{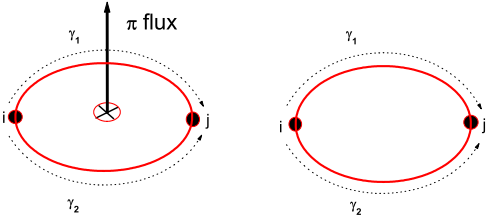}
\caption{The scheme of the interference on a torus. } \label{Fig.2}
\end{figure}

{For $Z_2$ topological orders on an $e*e$ lattice, the protected subspace
becomes two qubits.} To do arbitrary unitary transformation on the protected
subspace, one needs to apply the external field on a sub-lattice,
\begin{equation}
H^{\prime }=h\sum\limits_{i\in \mathrm{odd}}\sigma _i^x+h\sum\limits_{i\in
\mathrm{odd}}\sigma _i^y+h^z\sum\limits_i\sigma _i^z.
\end{equation}
This is because the perturbation $h\sum\limits_{i\in \mathrm{odd}}\sigma
_i^x+h\sum\limits_{i\in \mathrm{odd}}\sigma _i^y$ drives only Z2 vortex
hopping without affecting $Z_2$ charge. Hence the nine parameters of the
effective pseudo-spin model become $J_{xx}=g(\frac hg)^L$, $%
J_{yy}=J_{zx}=J_{xz}=0$, $J_{zz}=g(\frac{h_z}{2g})^{L_x+L_y},$ $\tilde{h}%
_1^x=g(\frac hg)^{L_x},$ $\tilde{h}_2^x=g(\frac hg)^{L_y},$ $\tilde{h}_1^z=g(%
\frac{h^z}{2g})^{L_y}$ and $\tilde{h}_2^z=g(\frac{h^z}{2g})^{L_x}.${\
Furthermore, on a manifold with higher genus, the degenerate ground states
become multi-qubit, which can be mapped onto multi-pseudo-spin model. By the
same method one can do TQC on the multi-qubit. }

\textit{Realization of the Wen-plaquette model }: Finally we discuss the
realization of the Wen-plaquette model in a optical lattice of cold atoms.
Because the Wen-plaquette model can be regarded as an effective model of the
Kitaev model on a two dimensional hexagonal lattice, one may realize the
Kitaev model firstly. The Hamiltonian of the Kitaev model is
\begin{equation}
\mathrm{H}=\sum_{j+l=\text{even}}(\mathrm{J}_x\sigma _{j,l}^x\sigma
_{j+1,l}^x+\mathrm{J}_y\sigma _{j-1,l}^y\sigma _{j,l}^y+\mathrm{J}_z\sigma
_{j,l}^z\sigma _{j,l+1}^z)
\end{equation}
where $j$ and $l$ denote the column and row indices of the lattice. In the
limit $\mathrm{J}_x\mathrm{\gg J}_z\sim \mathrm{J}_y$ of Kitaev model, the
effective Hamiltonian of the Kitaev model is simplified into the
Wen-plaquette model
\begin{equation}
H_0=-\frac{\mathrm{J}_z^2\mathrm{J}_y^2}{16|\mathrm{J}_x|^3}\sum_i\sigma _{%
\text{left}(i)}^x\sigma _{\text{right}(i)}^x\sigma _{\text{up}(i)}^y\sigma _{%
\text{down}(i)}^y.  \label{fw}
\end{equation}
Then one can use the Kitaev model in the limit $\mathrm{J}_x\mathrm{\gg J}%
_z\sim \mathrm{J}_y$ on a torus to do the TQC. The realization of
the Kitaev model on a two dimensional hexagonal lattice has been
proposed in Ref.\cite {du,zoller,zo1,gao,vid}. The essential idea
realizing the Kitaev model is to induce and control virtual
spin-dependent tunneling between neighboring atoms in the lattice
that results in a controllable Heisenberg exchange interaction.

\textit{Summary and discussion}: By controlling the tunneling
processes, one can do unitary transformation on a topological
protected qubit (the toric code), as paves a new road towards TQC
rather than by anyon-braiding. By using a designer Hamiltonian - the
Wen-Plaquette model as an example, we show how to control the toric
code to realize TQC. In particular, we give a proposal to the
measurement.

{In the end we give a short comment on the advantage and the disadvantage of
TQC by tuning quantum tunneling effect. For TQC by anyon-braiding, one needs
to} manipulate single quasi-particle which demands new techniques. On the
contrary, for the TQC by tuning tunneling, one needs only adjust the global
field strength (or direction). However, it is a true challenge to realize
the designed spin model on a manifold of higher genus in the optical lattice
of cold atoms. Such unsolved issue will be worked out in the future.

This research is supported by NFSC Grant no. 10574014. S.P. The author would
like to thank D. L. Zhou, B. Zeng and W.-B. Gao for helpful conversations.
The author is grateful to J. Vidal for kindly advices about \emph{%
controllable topological orders} which helped me to clarify some points in
this paper.


\begin{thebibliography}{99}
\bibitem{k1}  A. Kitaev, Ann. Phys. \textbf{303}, 2(2003).

\bibitem{k2}  A. Kitaev, Ann. Phys. \textbf{321}, 2(2006).

\bibitem{wen}  {X.-G. Wen}, \emph{Quantum Field Theory of Many-Body Systems}%
, (Oxford Univ. Press, Oxford, 2004).

\bibitem{ioffe}  L. B. Ioffe, M. V. Feigel'man, A. Ioselevich, D. Ivanov, M.
Troyer, G. Blatter, Nature 415, 503 (2002).

\bibitem{sa1}  S. Das Sarma, M. Freedman, C. Nayak, S. H. Simon, A. Stern,
arXiv: cond-mat/0707.1889.

\bibitem{pa}  J. K. Pachos, Annals of Physics 322, 1254 (2007).

\bibitem{sarma}  C. W. Zhang, S. Tewari, and S. Das Sarma, Phys. Rev. Lett.
99, 220502 (2007). S. Tewari, S. Das Sarma, C. Nayak, C. W. Zhang, and P.
Zoller, Phys. Rev. Lett. \textbf{98}, 010506 (2007).

\bibitem{du1}  Y.-J. Han, R. Raussendorf, and L.-M. Duan, Phys. Rev. Lett.
98, 150404 (2007).

\bibitem{zoller}  L. Jiang et al., Nature Phys., doi:10.1038/nphys943 (2008)
doi:10.1038/nphys943.

\bibitem{vid}  S. Dusuel, K.P. Schmidt, and J. Vidal, Phys. Rev. Lett.
\textbf{100}, 057208 (2008); Phys. Rev. Lett. \textbf{100}, 177204 (2008).

\bibitem{zhang}  Chuanwei Zhang, V. W. Scarola, Sumanta Tewari, and S. Das
Sarma, Proc. Natl. Acad. Sci. U.S.A. 104, 18415 (2007).

\bibitem{zhang1}  Chuanwei Zhang, S. L. Rolston, and S. Das Sarma, Phys.
Rev. A 74, 042316 (2006). Chuanwei Zhang, V. W. Scarola, and S. Das Sarma,
Phys. Rev. A 76, 023605 (2007).

\bibitem{gao}  C. Y. Lu, W.-B. Gao, O. Guhne, , X.-Q. Zhou, Z.-B. Chen,
J.-W. Pan, arXiv : quant-ph/0710.0278.

\bibitem{wen1}  X. G. Wen, {Phys. Rev. Lett. }\textbf{90} (2), 016803 (2003).

\bibitem{wen2}  X. G. Wen, Phys. Rev. D \textbf{68}, 065003 (2003).

\bibitem{du}  L.-M. Duan, E. Demler, and M. D. Lukin, Phys. Rev. Lett. 91,
090402 (2003).

\bibitem{zo1}  A. Micheli, G. K. Brennen, P. Zoller, Nature Physics, 2, 341
(2006).

\bibitem{bas}  G. Baskaran, S. Mandal, R. Shankar, Phys. Rev. Lett. \textbf{%
98}, 247201 (2007).
\end{thebibliography}
\end{document}